# Development of methods of the Fractal Dimension estimation for the ecological data analysis

Jakub Jura,
Department of Instrumentation and Control Engineering
Czech Technical University in Prague, FME
Prague, Czech Republic
jakub.jura@fs.cvut.cz

Aleš Antonín Kuběna
Institute of Information Theory and Automation
Czech Academy of Sciences
Prague, Czech Republic
akub@vsup.cz
Jakub Jura,

Martina Mironovová
Department of Instrumentation and Control Engineering
Czech Technical University in Prague, FME
Prague, Czech Republic
martina.mironovova@fs.cvut.cz

**Abstract:** This paper deals with an estimating of the Fractal Dimension of a hydrometeorology variables like an Air temperature or humidity at a different sites in a landscape (and will be further evaluated from the land use point of view). Three algorithms and methods of an estimation of the Fractal Dimension of a hydrometeorology time series were developed. The first results indicate that developed methods are usable for the analysis of a hydrometeorology variables and for a testing of the relation with autoregulation functions of ecosystem.

## Introduction

Fractal object is an ideal entity and its nature is epistemological. On a real objects is possible to see a partial characteristics of an ideal fractal object (especially there is possible to see here any degree of fractality). The difference between fractal and euclidean solid lies in the property which is called self-similarity – repeating same/or similar theme/shape in a different scales. This property causes relatively high segmentation of the object. And this segmentation is possible to see again and again in higher enlargement of a scale. This principle shown Benoit Mandelbrot [1] and Lewis Fry Richardn on the example of measuring of coastline length.

When we solving the task how to properly show the collected hydrometheorological data (Figure 2) from the database [2], we observe fractal character of these data. The key question is, that the fractal properties of hydrometeorological data have a relation to any ecosystem functions (like autoregulation).

## Fractal dimension

If we come back to example of coastline length measuring, so there is need to discuss the relation between value of a scale (or better measuring stick size) and the length of a coastline. It is clear that for fractal objects the length of coastline increases with decreasing a size of measuring stick. The steepness of this relation represents the degree of fractality of an object. And after the formal transformation is based for **Fractal Dimension** estimation.

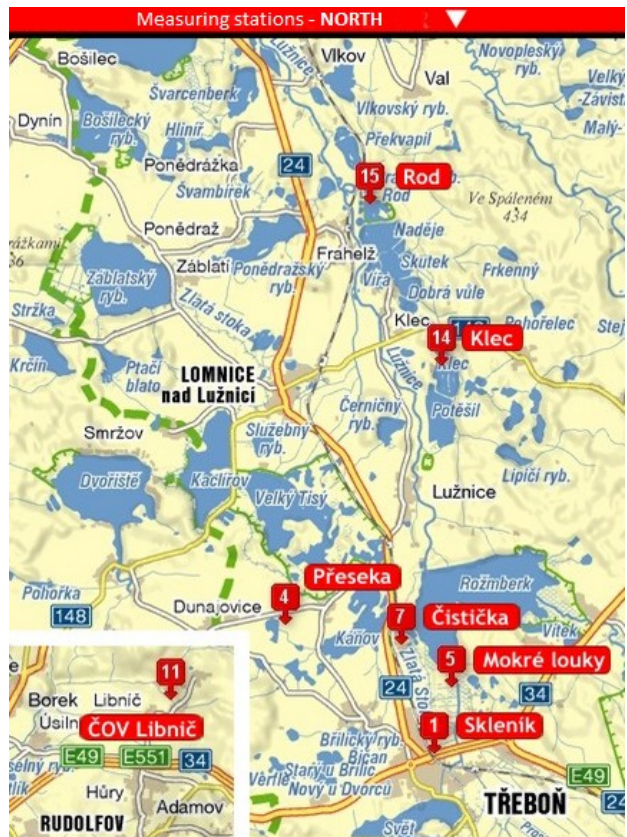

*Figure 1: Map of meteorological stations placement [1].*

There are many methods for Fractal dimension estimation. We are interested in a methods intended for the real fractal objects. Since the fractal theory is primarily intended for geometrical objects, so also Fractal Dimension estimation methods are also primarily intended to the geometrical objects (e.g. Box counting [3]). Unfortunately our fractal object is a time series and previous method is not possible to use directly.

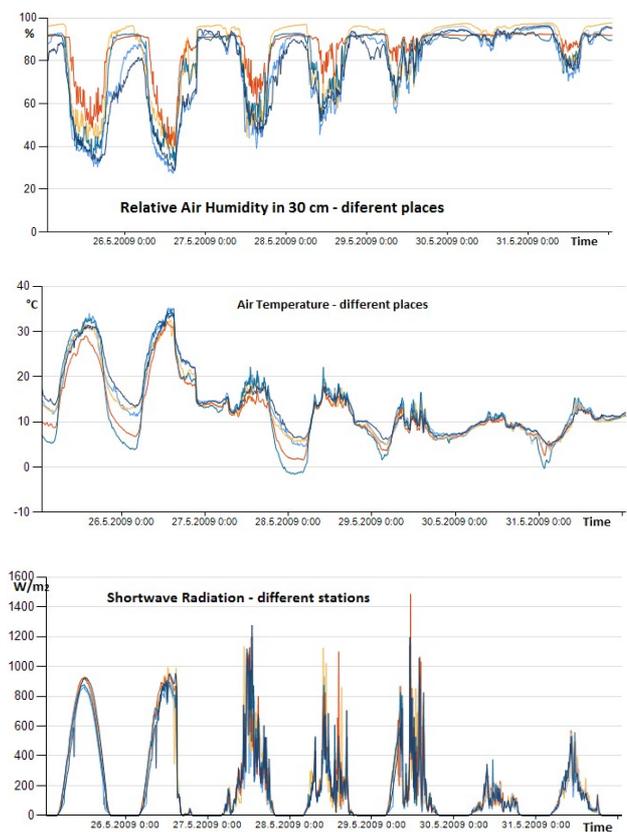

*Figure 2: Graph of measured data (Air humidity, temperature and incoming solar radiation) - one week range*

## Motivation

This research is continuation of the project Development of methods for evaluation of flows of energy and matters in the selected Eco-Systems and this project provided us a database of a hydrometheorological data which had collected from 16 meteorological measuring stations (Figure 3) in South Bohemia near Třebon town - Figure 1. Each station had measured a few basic hydrometheorological variables (Figure 2) like a Table 1:

| Air Temperature in 2 m and 30 cm abowe the ground | °C |
| --- | --- |
| Relative Air Humidity in 2 m and 30 cm abowe the ground | % |
| Incoming Solar Radiation | W/m² |
| Reflected Solar Radiation | W/m² |
| Precipitation | mm |
| Wind spead and wind direction | m/s, DEG |

*Table 1 Selected hydrometheorological variables from database TOKENELEK [2]*

Measuring stations were placed at a different typese of a sites {field, wet medow, concrete surface, fishpond, pasture, village} [4]. The aim of this article is to add a new point of view to the autoregulation in the landscape exploration. Especially the autoregulation of a temperature depends on the water in the landscape and on the state of a Small Water CycleSWC [5], [6], [7].

*Note: It is possible to estimate an autoregulation as an ability to regulate given environmental variable to a given value. E.g. the temperature is in the nature environment regulated to the temperature which is convenient to a local flora and fauna.*

The relation between diversity (especially biodiversity) of an ecosystem and a stability and the level of the autoregulation is currently discussed. To help to find the relation between fractal characteristics of a hydrometheorological variables and stability and the level of the autoregulation is the aim of this work. Partial aim is to develop the reliable method for estimation of Fractal Dimension of mentioned hydrometheorological variables.

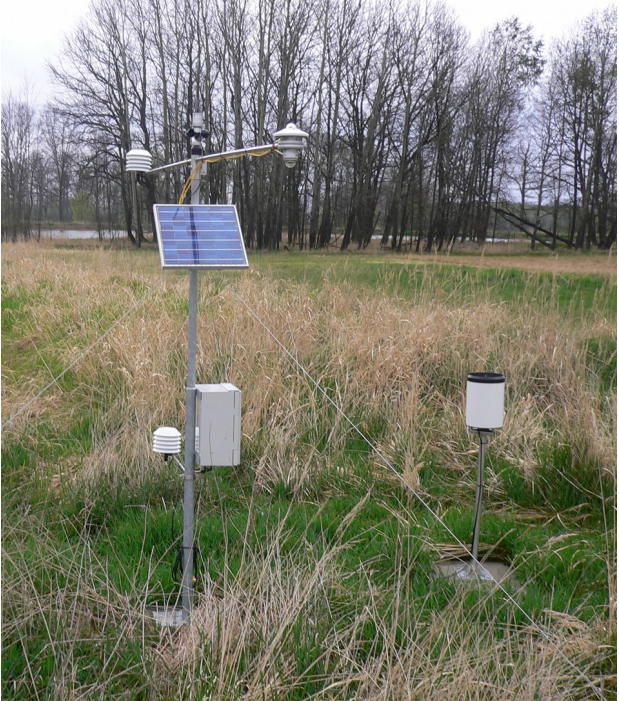

*Figure 3: One of meteorological station in South Bohemia [1]*

## Fractal dimension of time series

In general the idea of calculating of the fractal dimension is derived from a costline measuring and its general formulated [8]:

$$D = \frac{\log N}{\log(1/r)} \quad (1)$$

Where:
D … the fractal dimension
N … the number of an object's internal homotheties
r … the ratio of homothety

One of methods of the estimation of the Fractal dimension of real object is the Box Counting Method (e.g. [3], [9], [10]). This method is based on the covering of the object by the n-dimensional spheres, or cubes of given size ε.

$$D_C = \lim_{\varepsilon \to 0} \frac{\log[C(\varepsilon)]}{\log \frac{1}{\varepsilon}} \quad (2)$$

Where:
D … the estimation of FD
ε … the size of the n-dimensional cubes
C(ε) … the minimal number of n-dimensional cubes

But this method is suitable primarily for the geometrical object – not for time series (because induce an additional problem with choice of time-size scale). This method is possible to adapt to time series [11], but we move away from the original principle of Fractal dimension by use this method. Our aim is to use here methods which will closely corresponds to the principle of a coastline length measuring.

## Proposed methods of fractal dimension estimation

Here proposed methods is based on the principle of smoothing of the time series – and changing the coefficient of the smoothing. From the dependency between the length of the curve and smoothing coefficient is the value of Fractal dimension estimated. The primarily smooth course have a low Fractal Dimension estimation, because we smoothing of the smooth object. And the fractal dimension of a linear course is equal to zero. Next is description of three smoothing methods used for estimation of Fractal dimension – Moving Average, Moving Maximum and Minimum and method of Local Regression.

## Moving average (FDMAvg)

This method is based on the analogy with length of the coastline measuring. Similarly as a measuring stick size is changed (or box size ε in the case of Box counting method), so the window of the moving average is changed here.

**Description of algorithm**

Firstly the set of Window sizes E = {1,2,3,5,10,20, ...} is determined. Given window size is labelled here in according to Box Counting as $\varepsilon_n$, or $EPS_n$ and is chosen from this set.

For each data row index "k" and data x(k), from X the value AVG(k) of mean with given window size $\varepsilon_n$ is calculated (for k from 1 to m- $\varepsilon_{max}$, where m is inex of last element).

$$AVG(k) = \frac{1}{\varepsilon} \sum_{i=k}^{k+\varepsilon_n} x_i \quad (3)$$

Next is calculated the „variance "VAR".
Def: variance VAR is for our purpose the absolute value of a difference between two consequential values of AVG(k).
Continue is calculated the mean value of the variance for each window size $\varepsilon_n$ which is labelled VV_AVG(n):

$$VV\_AVG(n) = \frac{1}{m-\varepsilon_{max}} \sum_{k=1}^{m-\varepsilon_{max}} \left| \frac{1}{\varepsilon} \sum_{i=k}^{k+\varepsilon} x_i - \frac{1}{\varepsilon} \sum_{i=k}^{k+\varepsilon} x_{i+1} \right| \quad (4)$$

The estimation of the Fractal Dimension is the slope of the line approximating the dependency Log(VV_AVG) on Log(1/ε) - Figure 4.

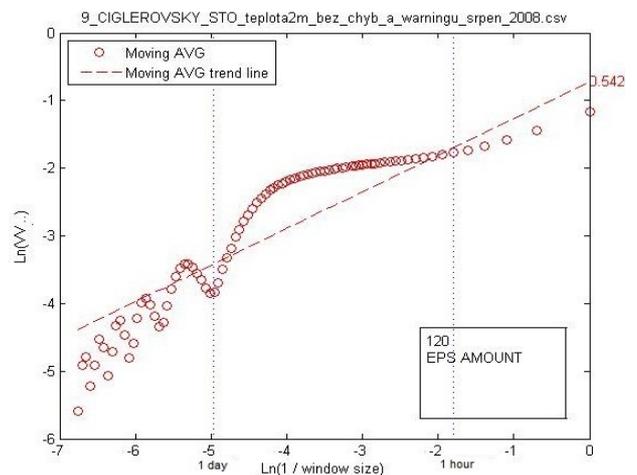

*Figure 4: dependency Log(VV_AVG) on Log(1/ε)*

**Methodology of use FDMAvg algorithm**
This method (similarly as others methods for estimation of Fractal Dimension) is sensitive to parameters of the calculation. For reach out the reliable results it is need to keep the methodological recommendation.

**1. Determination of minimal window size ($\varepsilon_{min}$)**
Since our data are sampled with the period 10 minutes, so we can use as a minimal window size the wide of one sample. The effect of the dynamical characteristic of the measuring instrument is insignificant. Hypothetically, in the case of a smaller sample period than time constant of measuring instrument, the result of fractal dimension estimation will be related more to the measuring instrument than measured object.

**2. Determination of maximal window size ($\varepsilon_{max}$)**
The determination of the maximum window size is not equally clear as a minimum windows size. For example Felix Hausdorf works with 1/5 of all range. Although a time series really has a finite length, theoretically can same series be longer (or newer end). Moreover the Fractal dimension estimation is significantly affected by this. Because basic data set has a length one month, so we work with maximal window size 1/5 of month.

**3. Step of the window size changes**
Because the method is based on a calculation of the slope of the line. And the line parameters are estimated by the minimal square method, so we have to ensure a uniform step of window size ε in logarithmic scale. In the opposite case the impact of the segment of points which are more close to a small values is significantly bigger than an impact of outlying ones.
Total amount of used window size is primarily not important for calculation (the line parameters influences minimally). A higher density of ε is useful for a visual evaluation of the results. For example on the Figure 5 is possible to see the step (close to one day ε size) and many waves in higher values. For a higher reliability of a comparison of many calculations is advantageous use same set of ε for all of them.

**Moving min&max (FDMM)**
This method is very similar as a previous – Moving Average FDMAvg. For analysis is used sums of moving minimums VV_MIN, sums of moving maximums VV_MAX and above all their mean value VV_MM. VV_MIN and VV_MAX have usually similar course. Bigger difference is possible to observe for sharply unsymmetric variables, like a precipitation. Normally the difference is small and we can estimate the Fractal dimension from the VV_MM (blue line on Figure 5).

**Description of algorithm**
The window sizes set is same as for method FDMAvg and also other procedures. The minimums MIN(k) and maximums MAX(k) are calculated instead of mean value AVG(k).

$$MIN(k) = \min(x_i); MAX(k) = \max(x_i)$$

for (5)

$$i = \{k,...,k+\varepsilon\}$$

Next is calculated the „variance" VAR for all rows. And the mean value of these variances VV_MIN(n) and VV_MAX(n) for each window size $\varepsilon_n$:

$$VV\_MIN(n) = \frac{1}{m-\varepsilon_{max}} \sum_{k=1}^{m-\varepsilon_{max}} |MIN(k) - MIN(k+1)|$$

$$VV\_MAX(n) = \frac{1}{m-\varepsilon_{max}} \sum_{k=1}^{m-\varepsilon_{max}} |MAX(k) - MAX(k+1)|$$

$$VV\_MM(n) = \frac{1}{2}(VV\_MAX(n) + VV\_MIN(n))$$

(6);(7);(8)

And finaly the estimation of the Fractal Dimension is again derive from the slope of the line approximating the dependency Log(VV_MM) on Log(1/ε). (Or VV_MAX or VV_MIN).

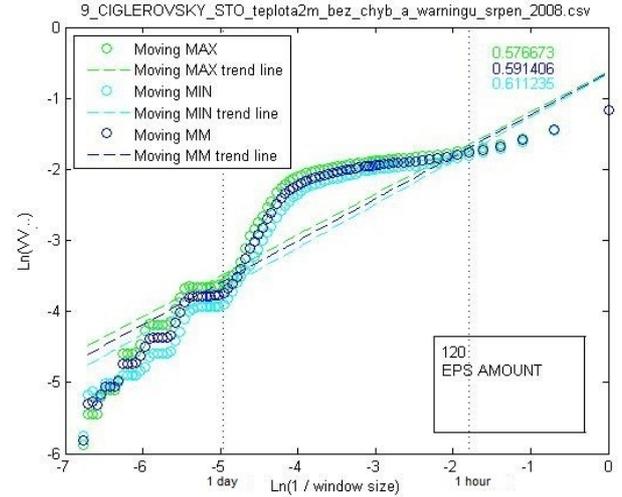

*Figure 5: dependency Log(VV_MAX) on Log(1/ε); Log(VV_MIN) on Log(1/ε); Log(VV_MM) on Log(1/ε) ... mean value of max and min.*

**Local regression (FDLR)**
For the parametrised smothing of the time series was used the smoothing spline method, which is special type of the local regression loes curve [12]. The aim of this method is to aproximate the data set of observations [$t_i$, $x_i$] by the function $y=y(t):[\min(t_i) \max(t_i)] \to R$, which is compromise between accuracy and smoothness. For the selected diferential operator $L: C^3 \to C^3$ and weight ratio p

$$RSS_p(y) = p\sum_i (y(t_i) - x_i)^2 + (1-p)\int_{t_{min}}^{t_{max}} L^2(y)dt \qquad (9)$$

We use L(y) = y''

$$RSS_p(y) = p\sum_i (y(t_i) - x_i)^2 + (1-p)\int_{t_{min}}^{t_{max}} (y'')^2 dt \qquad (10)$$

And approximation and substitution

$$\frac{d^2y}{dt^2} \approx \frac{y(t+\Delta) - 2y(t) + y(t-\Delta)}{\Delta^2} = \frac{(y_{n+1} - 2y_1 + y_{n-1})^2}{\Delta^2} \qquad (11)$$

Where selection of Δ has fulfill the condition of equidistant partition of the interval by the step Δ which cover all unobserved data.

If $\quad \lambda = \frac{1-p}{p}\Delta^{-3} \quad$ so we solve the optimisation task:

$$y = \arg\min_y \left( \sum_{n \in S} (y_n - x_n)^2 + \lambda \sum_n (y_{n+1} - 2y_1 + y_{n-1})^2 \right) \qquad (12)$$

$\lambda \in (0, \infty)$

S is set of indexes for which an observations exist. If λ → ∞ than y → at+b (optimum is close to linear regressive approximation of data. Conversely, for λ → 0, the output function y approximates measured date accurately.
For any λ (0, ∞), the optimisation task is convex and the sole local and global optimum is solution of the system of equations:

$$\frac{\partial}{\partial y_n}\left( \sum_{n \in S}(y_n - x_n)^2 + \lambda \sum_n (y_{n+1} - 2y_1 + y_{n-1})^2 \right) = 0 \qquad (13)$$

or better:

$$(F + \lambda E) = \hat{x}$$

where:

$$E = \begin{pmatrix} 1 & 2 & 1 & & & & \\ -2 & 5 & -4 & 1 & & & \\ 1 & -4 & 6 & -4 & 1 & & \\ & 1 & -4 & 6 & -4 & 1 & \\ & & ... & ... & 5 & -2 & \\ & & & & 1 & 2 & 1 \end{pmatrix} \qquad (14)$$

$F = diag(f)$

Where f = 1 for indexes for which the observation exist and f = 0 for indexes for which the observation does not exist. Vector $\bar{x}$ contains values of $x_i$ for indexes i for which the observation exist; and $x_i$ = 0 for i out of set S contains approximation at the point without measuring. Linear task was solved in Matlab software with help of the Sparse Matrix procedures.

This method provides us the matrix of smooth data which were used in similar way as a previous. The variance VV_LR was calculated, $\varepsilon_n$ was replaced by the $p_n$ and the Fractal Dimension was estimated from the given dependency Figure 6.

$$VV\_LR = |M(i,j) - M(i+1,j)| \qquad (15)$$

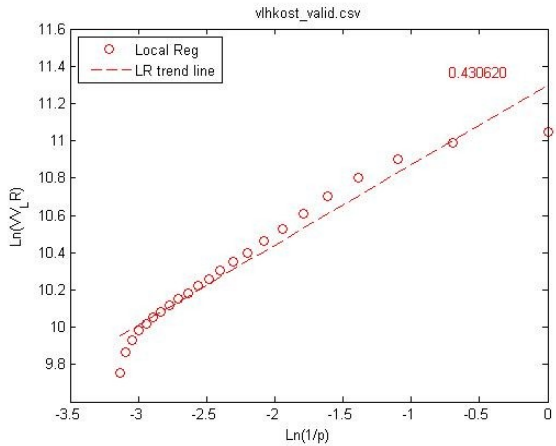

*Figure 6: dependency Log(VV_LR) on Log(1/ε);*

Also for this method is need to define the range of $\varepsilon_n$, especially the $p_n$ and situation is differ from previous because the $p_n$ parameter has no time dimension. The dependency on the parameter p from -32 to 32 has three parts:

    1. The first part is constant – there is not smoothing, because fineness of smoothing is smaller than sampling time of data.
    2. The second part decreasing linearly an is used for Fractal Dimension estimation and is shown on the Figure 6.
    3. The third part goes directly to zero.

The differences in Estimated Fractal Dimension by FDLR method between various hydrometeorological variables, corresponds relatively with results of FDAvg and FDMM.

**Preliminary results**

Before calculating has started, the database records had to been checked to an operational errors (in the database had been records which had represented breakdowns states of a measuring device). Moreover the basic analysis of data credibility had been made also.

The series of the first estimation was done for a few selected variables on a different sites – Figure 7. Here is possible to see that the Fractal Dimension of a temperature is higher at the locality with higher evapotranspiration (green areas versus concrete area or lake).

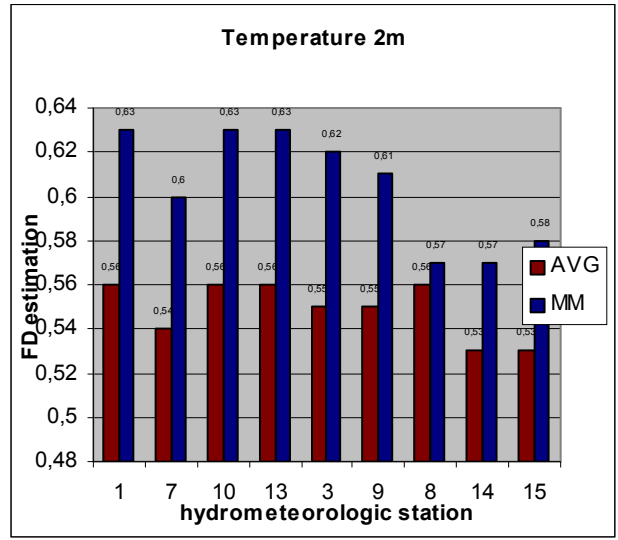

*Figure 7: Estimation Fractal Dimension of Temperature in 2m above the ground. Station 1 is in a small town, stations 7 has concrete surface, stations 10,13 are from medow, stations 3 and 9 are from wet terain and stations 8,14 and 15 from lake.*

This consideration is also supported by the example of a results of a Fractal Dimension estimation of Temperatures which is measured in 2 metres and 30 centimetres above the ground (Figure 8). The Fractal Dimension of variables measured more close to the ground points to influence of a Small Water Cycle. At the other side, the rigorous verification of this conclusion is not aim of this article and should to be done with consideration of many aspects of microclimatology.

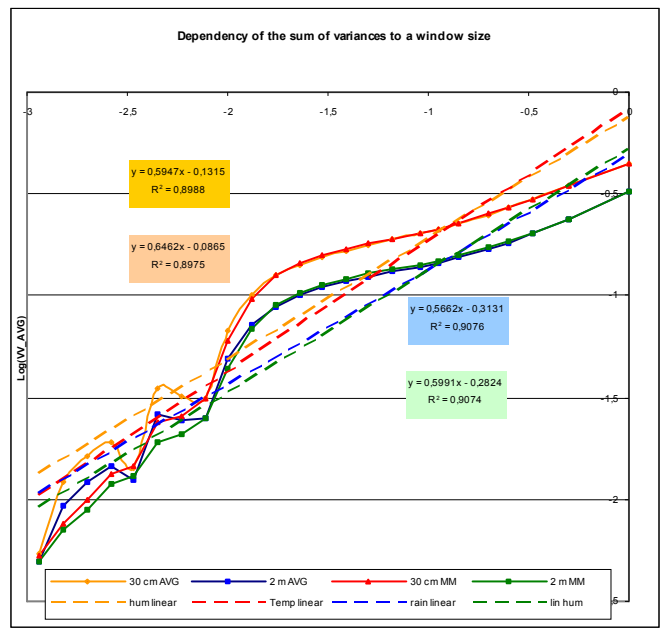

*Figure 8: Estimation of Fractal Dimension of temperatures at 30 cm and 2 m above the ground -station 9 CIGLEROVSKY_STO_All_Year_2008*

Also the differences between various hydrometeorological variables are observable at the level of a Fractal Dimension and also at the level of a course of dependency shown at Figure 9.

Question is what represent a waves between ε=((-2)...(-3)) in a mentioned dependency (shown at Figure 9) if we measure physically connected variables like a: air temperature, relative air humidity and incoming solar radiation. For example the spectral analysis would give an explanation of this phenomenon.

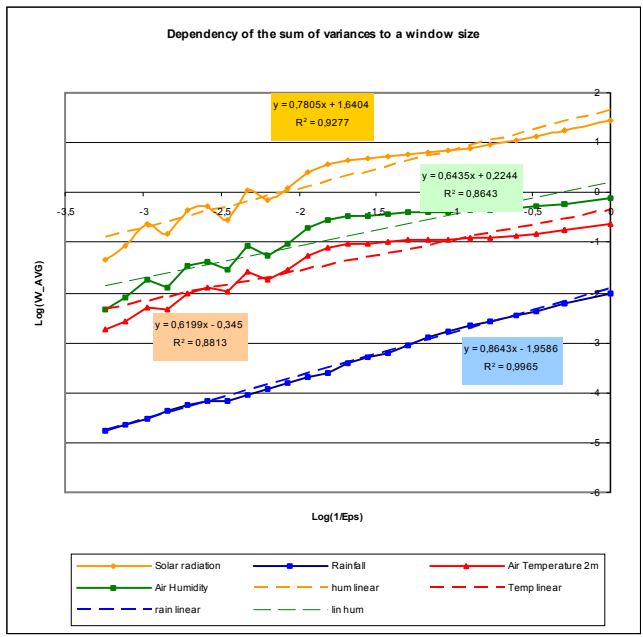

*Figure 9: Estimation of Fractal Dimension of different variables from station 9 CIGLEROVSKY_STO_july_2008, FDMAvg method.*

## Conclusion

In this article is developed three methods for an estimating of the Fractal Dimension of a hydrometeorology variables like an Air temperature, relative air humidity, incoming and reflected solar radiation, precipitation, soil humidity etc. at a different sites in a South Bohemia landscape. Mentioned method of Fractal Dimension estimation are based on relation between smoothed length of the course of the given hydrometeorological variable and independent smoothing parameter. For smoothing are used the moving average (and moving maximum and minimum) and local regression methods. All methods are developed at the level of algorithm and also at the level of methods, containing special parameters setting.

The preliminary results indicates that developed methods are usable for the analysis of a hydrometeorology variables and for a testing of the relation with autoregulation functions of ecosystem.

## Acknowledgements

This work was supported by the Excelence project P402/12/G097 DYME Dynamic Models in Economics of GACR.